\documentclass[final,3p,11pt,authoryear]{elsarticle}

\usepackage[none]{hyphenat}
\usepackage{amsmath,amssymb,amsthm}
\usepackage{natbib}
\usepackage{multirow}
\usepackage{graphicx} 
\usepackage[bookmarks, colorlinks]{hyperref} 
\hypersetup{linkcolor=blue,citecolor=blue,filecolor=black,urlcolor=blue} 
\usepackage{color}
\usepackage{geometry}

\begin{document}

\begin{frontmatter}

\title{Markov Switching}
\author[um]{Yong Song}

\author[um]{Tomasz Wo\'zniak\corref{cor1}} 

\cortext[cor1]{
\textit{Contact details:} Song: \href{mailto:yong.song@unimelb.edu.au}{yong.song@unimelb.edu.au}, Wo\'zniak: \href{mailto:tomasz.wozniak@unimelb.edu.au}{tomasz.wozniak@unimelb.edu.au}\\
\\
Article prepared for the Oxford Research Encyclopedia of Economics and Finance.\\
\\
The authors are grateful to Bill Griffiths, Chenghan Hou, Zhuo Li, Vance Martin, and Qiao Yang for their useful comments that improved the quality of this article.\\
\\
\textcopyright{} 2020 Yong Song \& Tomasz Wo\'zniak}
\address[um]{Department of Economics, University of Melbourne}

\begin{abstract}
Markov switching models are a popular family of models that introduces time-variation in the parameters in the form of their state- or regime-specific values. Importantly, this time-variation is governed by a discrete-valued latent stochastic process with limited memory. More specifically, the current value of the state indicator is determined only by the value of the state indicator from the previous period, thus the Markov property, and the transition matrix. The latter characterizes the properties of the Markov process by determining with what probability each of the states can be visited next period, given the 
state in the current period. This setup decides on the two main advantages of the Markov switching models. Namely, the estimation of the probability of state occurrences in each of the sample periods by using filtering and smoothing methods and the estimation of the state-specific parameters. These two features open the possibility for improved interpretations of the parameters associated with specific regimes combined with the corresponding regime probabilities, as well as for improved forecasting performance based on persistent regimes and parameters characterizing them.

The most commonly applied models from this family are the Markov switching ones that presume a finite number of regimes and exogeneity of the Markov process which is defined as its independence of the model unpredictable innovations. In many such applications, desired properties of the Markov switching model have been obtained either by imposing appropriate restrictions on transition probabilities or by introducing the time-dependence of these probabilities determined by explanatory variables or functions of the state indicator. One of the recent extensions of this basic specification includes Infinite Hidden Markov models that grant great flexibility and improved forecasting performance by allowing the number of states to go to infinity. Another one, namely, endogenous Markov switching model, explicitly relates the state indicator to model's innovations making it more interpretable and offering promising avenues for future developments.
\end{abstract}


\end{frontmatter}

\newpage
\section{Introduction}\label{sec:intro}

\noindent The Markov switching (MS) methodology was introduced by the seminal work of \cite{hamilton1989new}.
It is directly applicable to time series analysis for its dynamic nature. This section shows the benchmark model and corresponding notation for the data and model parameters. For a more comprehensive textbook exposition, see Chapter~22 by \cite{hamilton1994time}, \cite{Krolzig1997}, \cite{kim1999state} and \cite{fruhwirth2006finite}.

The dependent variable at time $t$ is denoted by $y_t$ for $t=1,...,T$ and $T$ is the number of periods in the sample.
$y_t$ can be either a scalar, vector, or matrix.
The independent variable is denoted by $x_t$, where it can be a scalar, vector, or matrix.
$x_t$ does not need to have the same dimension as $y_t$, and it might include lagged values of $y$.

\subsection{Markov process}\label{ssec:markovprocess}

\noindent A latent state at time $t$ is unobservable to an econometrician and denoted by $s_t$.
It takes the value $k\in\{1, 2, ..., K\}$, where $K$ is a positive integer representing
the total number of states. Such a latent variable $s_t$ indicates in which state the current system is at time $t$. Hence, it can be called \emph{state indicator} or \emph{regime indicator}.
The name \emph{state} and \emph{regime} are often used interchangeably in the literature.

From the definition, the state indicator is a scalar. 
However, it generalizes any vector or matrix representation of states as long as the number of states is finite.
For example, assume a vector of states $z_t=(z_{1t}, z_{2t})$, in which $z_{1t}$ takes values from set $\{1,...,K_1\}$
and $z_{2t}$ from $\{1,...,K_2\}$.
Such a tuple $z_t$ can be simply collapsed to a scalar indicator $s_t$, which takes value of $1,...,K_1K_2$.
Each element in $s_t$ corresponds to a vector $z_t$.
Sometimes, a vector form of the state variable is preferred in specific applications. 
However, knowing a scalar representation is generally enough to learn the basic framework.

The dynamics of the state indicator is governed by a Markov process. The probability distribution of $s_t$ given the whole path $\{s_{t-1},s_{t-2},...,s_1\}$ only depends
on the most recent state $s_{t-1}$.
Define such a \emph{transition probability} as
\begin{align}
\Pr(s_{t}=j\mid s_{t-1}=i, s_{t-2},...,s_1)&= \Pr(s_{t}=j\mid s_{t-1}=i) =  p_{ij}, \label{sy:eq:p}
\end{align}
where $i,j=1,...,K$. Given that in period $t-1$ the process was in state $i$ the probability that the state will switch to state $j$ in period $t$ is equal to $p_{ij}$. 
A \emph{transtion matrix} organizes these transition probabilities in a $K\times K$ matrix
and is definded as
\begin{equation}\label{sy:eq:P}
P=[p_{ij}]_{K\times K},
\end{equation}
where $p_{ij}$ is the element on the $i$th row and $j$th column such that the elements in each of the rows of matrix $P$ sum to one.

A vector of unconditional state probabilities, denoted by $\pi\equiv E[s_t]$, is defined from 
\begin{equation}\label{eq:pPp}
P' \pi = \pi.
\end{equation}
The equation above indicates time-invariance of distribution $\pi$, that is, an iteration over one period performed by the premultiplication by the transition probability $P'$ does not change vector $\pi$.
The solution to equation \eqref{eq:pPp} for $\pi$ given $\imath_K'\pi=1$, where $\imath_K$ is a $K$-vector of ones, is given in \citeauthor{hamilton1994time}~(\citeyear{hamilton1994time}, Chapter~22) and expresses $\pi$ as a function of $P$. Define a $(K+1)\times K$ matrix $\mathbf{P}=\begin{bmatrix} I_K - P' \\ \imath_K' \end{bmatrix}$. Hamilton's solution for $\pi$ is given by the $(K+1)$th column of $(\mathbf{P}'\mathbf{P})^{-1}\mathbf{P}'$. 

The distribution of the initial state at $t=0$, denoted by $s_0$, is represented by the following $K$-vector
\begin{equation}
\pi_0 \equiv [\Pr(s_0=k)]_{K\times1}. \label{sy:eq:pi0}
\end{equation}
For an ergodic Markov process, 
the initial distribution can be simply set to the stationary distribution $\pi$.
If the Markov process is non-stationary, there typically exists theory-guided information for $\pi_0$.
For example, a change-point model requires that the initial state is in the first state,
namely, $s_0=1$.

\subsection{Measurement Equation}\label{ssec:measurement}

\noindent A measurement equation lays out the probability law of the observations and is given by
\begin{align}
  y_t & \sim F(x_t, s_t), \label{sy:eq:y}
\end{align}
where $F$ represents the distribution of $y_t$ conditional on
the observation $x_t$ and the latent state $s_t$.
There is no restriction on $F$. It can be a discrete, continuous, or mixture distribution
depending on the structure of the data $y_t$. 
For example, consider a time series of asset returns and assume that each state admits an autoregressive process of order one with Gaussian innovations. Then $F$ can be expressed as
\[F(x_t, s_t) \equiv N(\mu_{s_t}+\beta_{s_t} y_{t-1}, \sigma^2_{s_t}),\]
where $\mu_{s_t}+\beta_{s_t} y_{t-1}$ is the mean and $\sigma^2_{s_t}$ the variance of this conditional distribution given $s_t$ and $y_{t-1}$.
The same assumption also implies a regression form of equation~\eqref{sy:eq:y} given by
\begin{equation}\label{eq:ar}
y_t = \mu_{s_t} +\beta_{s_t} y_{t-1} + \sigma_{s_t}\epsilon_t,\quad \epsilon_t\sim N(0, 1).
\end{equation}

Equation~\eqref{sy:eq:p} and \eqref{sy:eq:y} comprise the
foundation of the Markov switching framework.
Together with the initial condition \eqref{sy:eq:pi0},
the likelihood function $p(Y| \Theta)$
is available through the filter technique from \cite{hamilton1989new} called the \emph{Hamilton filter}.
The notation $Y$ denotes a collection of all $y_t$ for $t=1,...,T$;
and $\Theta$ is the collection of all time-invariant parameters.
For instance, $\Theta\equiv \{\mu_k, \beta_k, \sigma^2_k\}_{k=1}^K$ in the example above.

\subsection{Filtering the States}\label{ssec:filter}

\noindent The Hamilton filter gives the conditional distribution of the state $s_t$
on the data up to time $t$, $Y_{1:t}$, denoted by $p(s_t\mid Y_{1:t})$. It is used to compute the one-period ahead probability density of $y_t$
from the following formula.
\begin{subequations}
\begin{alignat}{6}
p(y_{t+1}\mid Y_{1:t}) & = \sum\limits_{s_{t+1}=1}^K p(y_{t+1}, s_{t+1}\mid Y_{1:t})\\
& =\sum\limits_{s_{t+1}=1}^K p(y_{t+1}\mid s_{t+1}, Y_{1:t})p(s_{t+1}\mid  Y_{1:t})\label{eq:likeli2}\\
& =\sum\limits_{s_{t+1}=1}^K \left[p(y_{t+1}\mid s_{t+1}, Y_{1:t})\sum\limits_{s_t=1}^K p(s_{t+1}, s_t\mid  Y_{1:t})\right]\\
& =\sum\limits_{s_{t+1}=1}^K \left[p(y_{t+1}\mid s_{t+1}, Y_{1:t})\sum\limits_{s_t=1}^K p(s_{t+1}\mid  s_t,  Y_{1:t})p(s_t\mid Y_{1:t})\right]\\
& =\sum\limits_{s_{t+1}=1}^K \left[p(y_{t+1}\mid s_{t+1}, Y_{1:t})\sum\limits_{s_t=1}^K p(s_{t+1}\mid  s_t)p(s_t\mid Y_{1:t})\right]\\
& =\sum\limits_{s_{t+1}=1}^K \sum\limits_{s_t=1}^K p(y_{t+1}\mid s_{t+1}, Y_{1:t}) p(s_{t+1}\mid  s_t)p(s_t\mid Y_{1:t}),
\end{alignat}
\end{subequations}
where $p(y_{t+1}| s_{t+1}, Y_{1:t})$ is obtained from the measurement equation~\eqref{sy:eq:y}
and $ p(s_{t+1}|s_t)$ is the transition probability in equation \eqref{sy:eq:p}. 
The derivation above utilizes the forecasted state probability conveniently decomposed into the transition and filtered probabilities
\begin{equation}\label{eq:forecasted}
p(s_{t+1}\mid  Y_{1:t}) = \sum\limits_{s_t=1}^K p(s_{t+1}\mid  s_t)p(s_t\mid Y_{1:t}),
\end{equation}
where the filtered probability for $s_{t+1}$ is given by 
\begin{align*}
p(s_{t+1}=k\mid Y_{1:t+1}) & = \frac{p(s_{t+1}, y_{t+1}\mid Y_{1:t})}{p(y_{t+1}\mid Y_{1:t})}\\
&=\frac{p(y_{t+1}\mid s_{t+1}, Y_{1:t}) p(s_{t+1}\mid Y_{1:t})}{p(y_{t+1}\mid Y_{1:t})}\\
&=\frac{p(y_{t+1}\mid s_{t+1}, Y_{1:t}) \sum\limits_{s_t=1}^K p(s_{t+1}\mid  s_t,  Y_{1:t})p(s_t\mid Y_{1:t})}{p(y_{t+1}\mid Y_{1:t})}.
\end{align*}
The filtered probability of $s_{t+1}$ is easy to compute as long as the filtered probability of $s_t$ is known.
The conditional density $p(y_{t+1}| Y_{1:t})$ has been derived above.
Given the initial distribution of the first-period state, $\pi_0$,
the filtered probability of $s_t$ and the associated one-period ahead probability density can be calculated iteratively.
The likelihood function is constructed as the product of conditional probability densities:
\begin{equation}\label{eq:likelihood}
p(Y\mid \Theta) = \prod\limits_{t=1}^T p(y_t\mid Y_{1:t-1}, \Theta),
\end{equation}
where $Y_{1:0}$ is treated as known and the state indicators, $s_t$ for $t=1,...,T$, are integrated out. An analytical expression for the likelihood function is available and can be found in \cite{fruhwirth2006finite}. It is based on equations \eqref{eq:likelihood} and \eqref{eq:likeli2}. 

\subsection{Parameter Estimation}\label{ssec:estimation}

\noindent The maximum likelihood estimation is based on an iterative expectation-maximization (EM) algorithm \citep[see][]{Hamilton1990}. In each of its iterations, a filtering-smoothing algorithm is used to propose the current estimate of $S$, a collection of all $s_t$ for $t=1,\dots,T$, and a maximization step is applied to compute an estimate of $\Theta$ and $P$. The maximum likelihood estimation is straightforward if regularity conditions are satisfied. However, for larger models, it might become cumbersome due to unbounded likelihood function. 

Bayesian estimation relies on a data augmentation technique that requires the specification of a complete-data likelihood function $p(Y,S|\Theta)=p(Y| S,\Theta)p(S|\Theta)$, where the objects on the right-hand side of the expression are easily obtainable without integration. The complete-data likelihood function is subsequently used to specify the full-conditional posterior distributions $p(\Theta| Y,S)$ and $p(S| Y,\Theta)$. Sampling from the joint posterior distribution of the parameters and states is performed via Markov Chain Monte Carlo (MCMC) methods. Since the former conditional distribution takes $S$ as given, it can be sampled from using standard techniques. The sampling from the latter relies on the \emph{forward filtering and backward sampling} (FFBS) method from \cite{chib1996calculating}. 

\subsection{Inference}\label{ssec:inference}

\noindent The first results for the asymptotic normality of the maximum likelihood estimator of the MS model parameters were proposed by \cite{Lindgren1978} and \cite{Lehmann1983}, whereas the finite sample properties of this estimator are analyzed in \cite{Psaradakis1998}. The frequentist solution to a problem of selection of the number of states of the Markov process $K$ relies on information criteria such as, for instance, Akaike Information Criterion \citep[see a detailed study by][]{Psaradakis2003}. Testing of a hypothesis that $K=1$ against an alternative hypothesis that $K=2$ is highly cumbersome since the MS model is not identified under the null hypothesis, and the solutions require sophisticated inferential methods some of which are provided by \cite{Carrasco2014} and more recently by \cite{Meitz2017}.

Bayesian model selection based on marginal data densities provides the solution to the number of states determination problem, including a model with $K=1$ \citep[see][]{fruhwirth2006finite}. However, the main challenge for correct Bayesian inference in MS models is the problem of label switching which is defined as the invariance of the likelihood function to the labeling of the states. Consider an example in equation~\eqref{eq:ar} with two states characterized by state-specific parameter vectors of state $A$ and state $B$, denoted respectively by $\Theta_A$ and $\Theta_B$.
The label switching problem states that irrespective of whether the vector of parameters $(\Theta_1,\Theta_2)$ is set to $(\Theta_A,\Theta_B)$ or $(\Theta_B,\Theta_A)$ the value of the likelihood function evaluated at these parameters stays invariant. \cite{Fruhwirth-Schnatter2001} proposes to analyze a multimodal global shape of the likelihood function and the posterior distribution \citep[see][for the application of this approach to the computation of marginal data densities]{Fruhwirth-Schnatter2004}. Implementing ordering restrictions on the state-dependent parameters of the model that would provide a unique classification of the states is a solution that is only applicable if such restrictions do not bind the posterior distribution. Alternatively, \cite{Geweke2007} proposes to base inference on label-switching invariant characteristics such as predictive densities.

\subsection{Interpretations}\label{ssec:interpretations}

\noindent The interpretation of the MS models is based on the combined analysis of the parameter estimates and one of the available estimates of the state probabilities. The latter should be chosen amongst the forecasted, filtered, or smoothed probabilities, denoted by $\Pr[s_t|y_{t-1}, y_{t-2}, \dots]$, $\Pr[s_t|y_{t},y_{t-1}, \dots]$ and $\Pr[s_t|Y]$, respectively, that are obtained through the FFBS algorithm, depending on a~particular application and the objective of investigation.

The first application of the MS models in economics included the analysis of business cycles and could be presented as follows. Consider data on gross domestic product growth rates to which the autoregressive model from equation \eqref{eq:ar} is fitted. The business cycle interpretation of the model was proposed by \cite{hamilton1989new}. It was acclaimed because the state-dependent constant term estimates were positive and negative, respectively, in the two states, which smoothed probabilities had high values for the periods commonly related to as economic expansions and recessions, respectively. Similar reasoning was applied to financial markets characterized by bull and bear markets subsequently occurring one after another as in \cite{Hamilton1996}. 

Another such example includes a multivariate model of the effects of monetary policy on the real economy in the U.S. with conditional heteroskedasticity modeled with the MS process proposed by \cite{Sims2006}. In this model, the subsequent volatility states occurred to have high values of smoothed probabilities in the periods corresponding to the terms of subsequent Chairs of the Federal Reserve. For instance, the state with the highest value of the volatility had high probabilities of occurrence in the period of Paul Volcker's chairmanship. In contrast, the lowest volatility state spanned the term of Alan Greenspan's.

Many applications in economics rely on some concept of causality. Two such popular concepts are Granger causality proposed by \cite{Granger1969} and \cite{Sims1980} that relates the causal link between variables to their predictive power, and another considers causal links between variables to be based on a structural model of an economy. An approach that investigates Granger causality for a specific state of an MS Vector Autoregressive model was proposed by \cite{Psaradakis2005} whereas the framework that is unconditional on the states was proposed by \cite{Droumaguet2017}. An explicit form of dependence between two or more Markov processes describing country or regional business cycles was proposed by \cite{Owyang2005}, \cite{Hamilton2012}, and more recently by \cite{Leiva-Leon2017}.

Finally, a new class of structural MS rational expectations models utilizes the MS rule to determine the time-variation of the parameters of the Dynamic Stochastic General Equilibrium model. In this framework, the considered agents are rational and, thus, know the MS rule and, consequently, take it into account in their decision-making problem. \cite{Farmer2009} provides the theory behind such a formulation of the model while \cite{Liu2011} provides a model for macroeconomic fluctuations. \cite{Farmer2011} propose a method of solving these models.

\section{Exogenous Markov Switching}\label{sec:exo}

\noindent The MS model was introduced in Section \ref{sec:intro} through the definition of the likelihood function in which the predictive densities of the data $p(y_{t+1}|s_{t+1},Y_{1:t})$ are weighted by the forecasted state probabilities $p(s_{t+1}|Y_{1:t})$   as in equation \eqref{eq:likeli2}. The decomposition of the latter into a product of the transition times the filtered probabilities as in equation \eqref{eq:forecasted} reveals that it does not depend on the contemporaneous observation $y_{t+1}$ but only on the previous state $s_t$ and past observations $Y_{1:t}$. The independence of forecasted state probability $p(s_{t+1}|Y_{1:t})$ from the error component of the measurement equation \eqref{sy:eq:y} defines a popular family of exogenous MS models with finite number of states, $K$, that is considered in this section. In Section \ref{sec:ihmm}, models with an infinite number of states, $K$, are considered while Section \ref{sec:endo}, discusses endogenous MS models in which this dependence is introduced.

\subsection{A Family of Markov Switching Models}\label{ssec:MS}

\noindent The properties of the latent Markov process are driven by the form of the transition matrix $P$. A general way of imposing restrictions on the transition matrix was proposed by \cite{Sims2008} and \cite{Wozniak2015c}. Let $P_i$ denote the $i$th row of matrix $P$ that is represented in terms of its unrestricted elements, collected in an $1\times r_i$ vector $p_i$ whose elements sum to one, as follows
\begin{equation}\label{eq:restrictP}
P_i = p_iW_i ,
\end{equation}
where $W_i$ are predetermined $r_i\times K$ matrices for $i=1,\dots,K$. The definition of the transition matrix as in the equation above is used below to demonstrate various types of this matrix and the implied properties of the Markov process. Note that if $r_i=1$ for some $i$, then the only element of $p_i$ is equal to one. 

A general stationary and aperiodic MS process in which each of the states can be revisited at any time $t$ presumes $r_i=K$ and $W_i=I_K$ for each $i$, where $I_K$ is an identity matrix of order $K$. In such a model, all of the $K^2$ elements of the transition matrix are estimated, there is no absorbing state, all of the elements of the ergodic probabilities vector $\pi$ are nonzero, and the probabilities of the initial state are most often set to $\pi$. This type of MS model is the most frequently applied in economics and finance. 

A class of finite mixture models is nested within the MS models by setting $W_i=I_K$ and $p_i=\pi'$ for each $i$, where all of the elements of $\pi$ are strictly positive. In this model, the forecasted state probabilities are time-invariant and equal to $p(s_{t+1}|Y_{1:t})=\pi$. However, the clustering of observations is facilitated through the smoothed probabilities that are allowed to change over time. Finite mixture models provide a convenient way of modeling nonstandard distributions that are often required for the error terms in economic and finance applications \citep[see][]{Norets2010}. It can be shown that any distribution of a random variable defined on a real scale can be approximated by a mixture of normal distributions, while distribution of a random variable defined on a positive real scale can be approximated by a mixture of gamma distributions.

Change-point models can be used to introduce monotonic regime changes as in a model proposed by \cite{Chib1998}. The process is initiated in the first state $s_0=1$, with probability $p_{11}$ it remains unchanged, and with probability $p_{21}$ it switches to the other regime. The first state is never to be revisited and, thus, $p_{12}=0$. In general, given that at some period $t$ the Markov process is in state $s_t=k<K$, it remains in this state with probability $p_{kk}$ and is only allowed to switch to the next regime with probability $p_{k+1.k}$. Finally, when the process reaches the $K$th state, it stays there forever. In the change-point models the transition matrix is obtained by setting $r_i=2$ and $W_i$ to a $2\times K$ matrix whose columns contain the $i$th and $(i+1)$th rows of $I_K$ for $i=1,\dots,K-1$, as well as $r_K=1$ and a row vector $W_K$ contains the last row of $I_K$. An example of such a transition matrix for $K=3$ is given by
$$ 
\begin{bmatrix}
p_{11}&p_{12}&0\\
0&p_{22}&p_{23}\\0&0&1
\end{bmatrix}.
$$
Therefore, these models are capable of estimating the time at which the regime changes occur. Consequently, they attracted considerable attention in economic and finance applications despite some recent evidence that maintaining the transition matrix unrestricted is capable of capturing similar patterns in data with non-deteriorating in- and out-of-sample fit. Finally, the change-point models introduce non-stationarity in the Markov-process and, thus, their ergodic probabilities are all equal to zero except for the last element of $\pi$ that is equal to one. Therefore, the last state is the absorbing state that gains 100 percent of the probability mass asymptotically with $T\rightarrow\infty$. Finally, \cite{fruhwirth2006finite} gives a detailed discussion of the nuances of the estimation of stationary and non-stationary Markov processes.

In a simple and popular deterministic change-point models, $s_t$ is assumed to be known and provided by the econometrician. In many applications, this model is used in order to estimate state-dependent parameters in pre-determined subsamples. Moreover, it is straightforward to set $s_t$ to obtain monotonic regime changes. Nevertheless, in deterministic change-point models, the properties of the Markov process are not driven by the transition matrix, which is redundant given fixed $s_t$. The regime-change dates are not estimated and, unless the econometrician knows the data generating process and sets $s_t$ accordingly, the model fit deteriorates heavily compared to other models considered in this section.

A more elaborate form of matrices $W_i$ may lead to the desired application-specific properties of the Markov process. Consider a model used by \cite{Sims2001} who introduces symmetric jumping among adjacent regimes. In this example, the desired transition matrix for the case of $K=4$ states is 
$$
\begin{bmatrix}
p_{11}&(1-p_{11})&0&0\\
(1-p_{22})/2&p_{22}&(1-p_{22})/2&0\\
0&(1-p_{33})/2&p_{33}&(1-p_{33})/2\\
0&0&(1-p_{44})&p_{44}
\end{bmatrix},
$$
which can be easily obtained by setting $r_i=2$ for each $i$ and
$$
W_1=\begin{bmatrix}
1&0&0&0\\0&1&0&0
\end{bmatrix}, \quad 
W_2=\begin{bmatrix}
0&1&0&0\\\frac{1}{2}&0\frac{1}{2}&0
\end{bmatrix}, \quad
W_3=\begin{bmatrix}
0&0&1&0\\0&\frac{1}{2}&0&\frac{1}{2}
\end{bmatrix}, \quad
W_4=\begin{bmatrix}
0&0&1&0\\0&0&0&1
\end{bmatrix}.
$$
In another example, \cite{Sims2008} consider a generalization of this model inspired by the developments in \cite{Cogley2005} that allows for potentially many states and the estimation of all of the parameters in matrix $P$, while keeping the number of its free parameters relatively small. In this model, the desired transition matrix has the following form
$$
\begin{bmatrix}
p_{11}&a_1\alpha_1(1-p_{11})&a_1\alpha_1^2(1-p_{11})&\cdots&a_1\alpha_1^{K-1}(1-p_{11})\\
a_2\alpha_2(1-p_{22})&p_{22}&a_2\alpha_2(1-p_{22})&\cdots&a_2\alpha_2^{K-2}(1-p_{22})\\
\vdots&\vdots&\vdots&\ddots&\vdots\\
a_K\alpha_K^{K-1}(1-p_{KK})&a_K\alpha_K^{K-2}(1-p_{KK})&a_K\alpha_K^{K-3}(1-p_{KK})&\cdots&p_{KK}\\
\end{bmatrix},
$$
that can be obtained by setting $r_i=2$ for each $i$, matrices $W_i$ accordingly to
\begin{align*}
W_1&=\begin{bmatrix}
1&0&0&\dots&0\\
0&a_1\alpha_1&a_1\alpha_1^2&\dots&a_1\alpha_1^{K-1}
\end{bmatrix}, \\
W_2&=\begin{bmatrix}
0&1&0&\dots&0\\
a_2\alpha_2&0&a_2\alpha_2&\dots&a_2\alpha_2^{K-2}
\end{bmatrix},\\
\dots \\
W_K&=\begin{bmatrix}
0&0&0&\dots&1\\
a_K\alpha_K^{K-1}&a_K\alpha_K^{K-2}&a_K\alpha_K^{K-3}&\dots&0
\end{bmatrix},
\end{align*}
where $\alpha_1,\dots,\alpha_K$ are positive free parameters to be estimated, and $a_1,\dots,a_K$ are such that the columns of matrices $W_i$ sum to one. This model uses scarce parameterization of the transition matrix that is capable of capturing occasional discontinuous shifts in the values of the regime-dependent parameters when $K$ is small, as well as frequent, incremental changes in these parameters for larger $K$. The choice of $K$ is a matter of empirical investigation.

Finally, the Markov property of the latent process might be extended by introducing the dependence of the current state, $s_t^*$, not only on one but several recent realizations of the latent process. The original model by \cite{hamilton1989new} assumes the dependence of the model parameters on the current and previous regime, $s_t^*$ and $s_{t-1}^*$ respectively. This dependence can be modeled by a new four-state Markov process, $s_t$, through the following state representation:
\begin{align*}
s_t = 1&\quad\text{ if } s_t^*=1 \text{ and } s_{t-1}^*=1,\\
s_t = 2&\quad\text{ if } s_t^*=2 \text{ and } s_{t-1}^*=1,\\
s_t = 3&\quad\text{ if } s_t^*=1 \text{ and } s_{t-1}^*=2,\\
s_t = 4&\quad\text{ if } s_t^*=2 \text{ and } s_{t-1}^*=2,
\end{align*}
and an appropriate form of the transition matrix:
$$
\begin{bmatrix}
p_{11}&p_{12}&0&0\\
0&0&p_{21}&p_{22}\\
p_{11}&p_{12}&0&0\\
0&0&p_{21}&p_{22}\\
\end{bmatrix},
$$
that can be specified by setting $r_1=r_2=2$, $p_1=p_3$ and $p_2=p_4$ and 
$$
W_1=W_3 = \begin{bmatrix} 1&0&0&0\\0&1&0&0 \end{bmatrix},\quad
W_2=W_4 = \begin{bmatrix} 0&0&1&0\\0&0&    0&1 \end{bmatrix}.
$$

\subsection{Independent Markov Processes}\label{ssec:heteroMS}

\noindent In this class of MS models, various groups of parameters of the model depend on separate and independent Markov processes \citep[see][for some early applications]{Phillips1991,Ravn1995}. The examples of such models include models in which the parameters of the conditional mean process depend on a different Markov process than conditional variances \citep[see, e.g.][]{Sims2008}, or structural models in which the money demand equation depends on different Markov process than other parameters of the models \citep[see, e.g.][]{Sims2006}. Consider $L$ such independent processes $s_{lt}$ each parameterized by a $K^l\times K^l$ transition matrix $P^l$ for $l=1,\dots,L$. This model can be represented by a composite Markov process $s_t=(s_{1t},\dots,s_{Lt})$ with $\prod_{l=1}^{L}K^l$ states and the corresponding transition matrix given by
$$
P = P^1\otimes\dots\otimes P^L,
$$
where $\otimes$ denotes the Kronecker product. The gain from the tensor product representation of the transition matrix above, introducing nonlinear restrictions, is an economic parameterization facilitating the estimation. Note that each of the transition matrices $P^l$ might be subject to restrictions as that in equation \eqref{eq:restrictP}.

\subsection{Non-homogeneous Markov Switching}\label{ssec:nonhomoMS}

\noindent Finally, this survey of parameterizations of transition matrices is concluded by presentation of a non-homogenous MS model in which the transition probabilities change over time \citep[see][]{Filardo1994,Diebold1994}. The introduction of this time variation is often combined with the dependence on some variables $v_t$ that might contain $x_t$ \citep[as in e.g.][]{Filardo1994} as well as the state indicators $s_t$ \citep[as in][]{Otranto2005} or a measure of a duration of the state \citep[as in][]{Durland1994}. The restriction imposed on the columns of the transition matrix leads to the parameterization of the transition probabilities through the logistic regression in the following form
\begin{equation*}
  p_{ij}\equiv P(s_{t}=j\mid s_{t-1}=i) = \frac{\exp(v_t\gamma_{ij})}{\sum_{j=1}^{K}\exp(v_t\gamma_{ij})},
\end{equation*}
where $\gamma_{ij}$ are parameter vectors to be estimated the dimensions of which correspond to vector $v_t$. Note that for the identification of transition probabilities for each $i$ there is a $j$ so that $\gamma_{ij}$ is a vector of zeros \citep[see][for a recent exposition on the setup of the model]{Koki2020}. The selection of the variables in $v_t$ determines the time-dependence in transition probabilities and is subject to empirical verification. As long as the establishment of the factors determining transition probabilities might be of interest in itself the non-homogeneous MS models often lead to similar estimates of filtered and smoothed state probabilities as the simple MS model in many data sets.

\section{Infinite Hidden Markov Model}\label{sec:ihmm}

\noindent The infinite hidden Markov model (IHMM) was developed by \cite{beal2002infinite} and \cite{teh2005hierarchical}.
It builds on the Dirichlet process mixture model (DPM) from \cite{escobar1995bayesian}
and extends the finite number of states of the MS model to the case in which this number goes to infinity, $K\rightarrow \infty$.
Such an extension introduces a fundamental advancement of econometric modeling 
by transforming the parametric MS framework into a nonparametric structure.

A direct consequence is that the transition matrix $P$ implied by equation \eqref{sy:eq:P} has an infinite dimension and can be presented as
\[P\equiv [p_{ij}]=
\begin{bmatrix}
  p_{11} & p_{12} & p_{13} & ... \\
  p_{21} & p_{22} & p_{23} & ... \\
  p_{31} & p_{32} & p_{33} & ... \\
  \vdots &  \vdots &  \vdots & \ddots \\
\end{bmatrix},\]
where $i,j=1,2,3,\dots$ and $p_{ij}\geq 0$.
From the definition, each row of $P$ must sum up to $1$, $\sum\limits_{j=1}^\infty p_{ij}=1$.
The time-invariant parameters that describe the $k$th state are defined as $\theta_k$, and there is an infinite number of them. 
Similarly to the finite-state MS models, the parameter space comprises of the state indicator $S\equiv \{s_t\}_{t=1}^T$,
the time-invariant parameters $\Theta\equiv \{\theta_k\}_{k=1}^\infty$
and the transition matrix $P=[p_{ij}]_{\infty\times \infty}$.

\subsection{Estimation}\label{ssec:estimationIHMM}

\noindent Because of the parameter saturation problem,
the IHMM cannot be estimated by classical methods without regularization.
On the contrary, the Bayesian approach is coherent, more appropriate for inference, and thus, the existing research on the IHMM uses mostly this framework.

Three popular ways could be used to draw inference from the IHMM.
The first is to integrate out the transition probability $P$
based on the Chinese restaurant representation of the Dirichlet process as in \cite{fox2011sticky}.
This method works directly on states, but the derivation is complicated.
The second is to apply the beam sampler to stochastically truncate the number of states
to a finite one during the MCMC as in \cite{van2008beam}.
This method provides an exact inference similarly to the first method. Besides,
it utilizes conditional independence by keeping the transition matrix 
$P$ in the parameter space, which allows parallel computations 
and is usually much faster than the first approach.

The last method applies the degree-$K$ weak limit approximation from \cite{Ishwaran02}
as in \cite{bauwens2017autoregressive}.
It uses a truncated Dirichlet process so that the IHMM resembles an appropriate finite-state MS model.
In practice, \cite{song2014modelling} found that standard MS models with a large number of states
performs similarly to the IHMM, where the number of inactive states should be nonzero, where an inactive state is a state without data being assigned to it.
This approach renders the IHMM easier to execute.
Although two caveats exist. First, the prior distribution on the transition matrix must be chosen according to \cite{Ishwaran02}.
Namely, the concentration parameter from the truncation approximation should be consistent with 
the concentration parameter from the IHMM.
Otherwise, the approximation is not valid.
Second, the number of states in the MCMC must be monitored to avoid poor approximation.
A simple rule would be that the number of active states, that is those with nonzero number of observations classified into it, must always be less that 
the total number states $K$ in the approximation.
For execution of the algorithm, see \cite{bauwens2017autoregressive}.

\subsection{Flexibility of IHMM}\label{ssec:flexible}

\noindent The IHMM originated from the machine learning literature where it was used to reveal dynamic 
clustering in applications, including dialogue summary and motion capture.
For the last decade, it has been applied to various fields in economics and finance. 
The seminal developments include \cite{song2014modelling}, 
\cite{jochmann2015modeling} and \cite{dufays2015infinite}.
The motivation for using the IHMM is that it demonstrates well the trade-off between heuristic economic interpretations and competitive forecast accuracy.
This feature constitutes an advantage over many of the machine learning methods that hardly allow for structural interpretations. 

Similarly to the conventional MS models, the IHMM maintains its first-order Markov chain property.
However, due to the differences in the prior distribution setup, and the problem of doubling the states, a phenomenon consisting of the possibility of producing an additional state that mimics an already existing one, the DPM should not be considered a device for detecting the number of states of the mixture or MS models as it was suggested in some early approaches such as by \cite{Otranto2002}. \cite{miller2013simple} states the argument formally.

An attractive feature of the IHMM is that it captures regime-switching and structural break jointly. Therefore it grants more flexibility and accommodates data dynamics upon the arrival of new observations. 
The regime-switching module pools data with similar behavior 
to borrow statistical strength from each other.
In addition, the structural break dynamics can generate a
new state when the arriving new observations exhibit a new law of motion.
A well-known example is the global financial crisis in 2007 and 2008.
It began with the subprime mortgage crisis in the U.S. and became the most severe financial crisis since 1930. Such an asset market downturn
does not have the same source as any bear market regime in history.
Any standard MS model for bull and bear markets
such as \cite{maheu2000identifying}, \cite{lunde2004duration}
and \cite{maheu2012components} is incapable of capturing the new phenomenon,
because they are limited by the data history and do not allow structural changes.
The IHMM is an appropriate vehicle to achieve such a goal because the capability of generating a new unprecedented state is a feature of the latent process.

Another advantage of the IHMM lies in its superior forecasting performance,
which is an empirical observation with plausible intuition.
Benchmark models such as autoregressions or linear regression have a rigid assumption 
of the functional forms.
Instead, the IHMM is more flexible and hence robust to model misspecification.
Its flexibility comes from its ability to sort data into clusters endogenously,
so that behaviorally different data will not pollute each other's inference.
One can find the kinship between this idea and the piece-wise linear model in a univariate framework.
Meanwhile, the IHMM is easily extendible to the multivariate framework, and the nodes are inferred jointly with the other model coefficients.
Two vital components to achieve better prediction performance are the hierarchical prior structure \citep{song2014modelling}
and certain regime persistence \citep{fox2011sticky} in economic and finance applications.
The hierarchical structure exploits more data information by letting the regimes to inform one another.
At the same time, regime persistence reflects salient stylized facts that economic time series are prone to
local dynamic stability.

The IHMM provides a convenient tool for a control approach in grand modeling frameworks.
Consider a modeling framework that utilizes independent variables $x_t$ and error term $\epsilon_t$.
Any incorrect distributional assumption about $\epsilon_t$ could potentially adversely affect the inference on the functional form of $x_t$.
If such distribution is not the focus of the application, simple estimators as (nonlinear) least-squares are perfectly competent.
If the distribution matters for inference such as in risk analysis,
these methods are no longer useful, and a semi-parametric approach has its advantages by imposing a nonparametric distribution on $\epsilon_t$.
Such an assumption releases $\epsilon_t$ from any potential misspecification.
Moreover, the parametric part related to $x_t$ is free from contamination of any parametric assumption imposed in $\epsilon_t$.
Lastly, the curse of dimensionality is not an issue because $\epsilon_t$ is usually univariate, while $x_t$ is not.
For examples, see \cite{jensen2010bayesian}, who modeled $\epsilon_t$ as a DPM model.
The most recent research treating $\epsilon_t$ as an IHMM includes \cite{hou2017infinite} and \cite{dufays2019sparse}.

\subsection{Extensions of IHMM}\label{ssec:extensions}

\noindent The IHMM provides a basis for the burgeoning academic literature on its extensions.
An IHMM with DPM emission can be found in \cite{fox2011sticky}, but it does not allow sharing particles among states.
A block IHMM by \cite{stepleton2009block} generalizes the stick structure of \cite{fox2011sticky}
by introducing more in-state dynamics to capture finer structures.
In this approach, each state is a distinct MS model, that makes it suitable in the application to the bull and bear markets modeling as in \cite{maheu2012components}.
To capture long memory, \cite{gael2009infinite} proposed the factorial IHMM.
Examples of its applications include \cite{nakano2011bayesian} and \cite{heller2009infinite}.
The factorial IHMM does not impose any restrictions for identification and, thus, structural interpretations are hardly possible, which limits the scope of applications in economics and finance. Many interesting modeling frameworks arise from incorporating various components into the IHMM structure, and some of such ideas can be found in \cite{gray1996modeling}, \cite{ehrmann2003regime} and \cite{haas2004new}. Alternatively, a modeling strategy consisting of applying the IHMM structure to existing interpretable parametric models can be found in \cite{liu2018improving}, \cite{jin2019bayesian} and \cite{jin2019bayesian}. Finally, there is a substantial body of work that follows up and extends the IHMM in different directions and some examples include \cite{shi2014identifying}, \cite{bauwens2017autoregressive}, \cite{maheu2016infinite}, \cite{yang2019stock}, \cite{hou2017infinite} and \cite{luo2019forecasting}.


\section{Endogenous Markov Switching}\label{sec:endo}

\noindent Recent literature proposed models that question the assumption of exogeneity of the Markov process and make an explicit link between the measurement equation error term and this latent process. In consequence, some form of endogeneity of the Markov process is introduced. \cite{Kim2008} argue that in many applications, endogeneity of the Markov process corresponds to the data properties and theoretical considerations to a larger extent than the exogenous one. While the initial proposal by \cite{Kim2008} implements endogeneity in the MS model, this section focuses on a specification proposed by \cite{Chang2017}. In this model, a discrete-valued latent process, $s_t$, is driven by a real-valued process, $w_t$, that is related to the measurement equation error term and then conveniently discretized. 

\subsection{Introducing Endogeneity}\label{ssec:introendo}

\noindent Define the dynamics of the real-valued latent factor by an autoregressive process
\begin{equation}\label{eq:rv-latent}
w_t = \alpha w_{t-1} + v_t,
\end{equation}
where $\alpha$ is the persistence coefficient such that $|\alpha|\leq1$ and $v_t$ is a standard normal error term. The initial value of the process, $w_0$, is recommended to be normally distributed with the zero mean and variance equal to $1/(1-\alpha^2)$ for $|\alpha|<1$, or equal to zero, $w_0=0$, for $\alpha=1$. The process in equation \eqref{eq:rv-latent} is discretized by defining a threshold parameter, $\tau$, and the discrete-valued state indicator for a two-state model, $K=2$, as
\begin{equation}\label{eq:dv-latent}
s_t = \left\{\begin{array}{ll} 1 & \text{ for } w_t<\tau, \\ 2 & \text{ for } w_t\geq\tau.  \end{array} \right.
\end{equation}
Therefore, as long as the process $w_t$ is subject to interpretation, its primary role is to define the Markov process $s_t$. 

\cite{Chang2017} define the measurement equation in a general form that makes an explicit link between the potential dependence of the conditional mean and standard deviation on the independent variables, $x_t$, and the latent processes
\begin{equation}\label{eq:endo-measurement}
y_t = m(x_t, w_t) + \sigma(x_t, w_t)\epsilon_t = m(x_t, s_t) + \sigma(x_t, s_t)\epsilon_t,
\end{equation}
where $\epsilon_t$, conditionally on $x_t$ and $w_t$ (or $s_t$), is a serially uncorrelated standard normal error term. Endogeneity of the Markov process is formally introduced in this model by an appropriate specification of the joint distribution of the error terms from the state and measurement equations \eqref{eq:rv-latent} and \eqref{eq:endo-measurement} respectively that is given by
\begin{equation}\label{eq:endo-error}
\begin{bmatrix} \epsilon_t \\ v_{t+1} \end{bmatrix} \sim \mathcal{N}\left( \begin{bmatrix}0\\0 \end{bmatrix}, \begin{bmatrix}1 & \rho\\ \rho & 1 \end{bmatrix} \right),
\end{equation}
where $\rho$ is the correlation coefficient. It can be shown that the exogenous MS model that is discussed in Section \ref{sec:exo} can be obtained by setting $\rho=0$. In other words, the restricted endogenous and the exogenous MS models are observationally equivalent, that is, they both lead to the same value of the likelihood function given the values of $\Theta$ and $S$. However, for the values of $\rho$ that are different from zero the relationship between $\epsilon_t$ and $w_{t+1}$ and, consequently, between $\epsilon_t$ and $s_{t+1}$ is established. Expressions for the implied transition probabilities that in this model are changing over time are given in the original paper.

A different form of endogeneity was proposed by \cite{Kim2008} who specified the joint distribution of error terms, such as the one in equation \eqref{eq:endo-error}, for vector $(\epsilon_t,v_t)$. This model presumes a contemporaneous effect of $\epsilon_t$ on $w_t$ and $s_t$ which \cite{Chang2017a} point out to be misspecified. \cite{Kim2008} consider an application to modeling a volatility feedback effect in financial time series.

\subsection{Interpretations Considering Correlation Coefficient}\label{ssec:interpret}

\noindent To illustrate the interpretation of non-zero $\rho$ consider a model in equation \eqref{eq:ar} with $s_t$ specified by the endogenous Markov process and with $|\beta_k|<1$ for $k=1,2$. A possible application in finance includes the modeling of the leverage effect defined as the negative correlation between the current innovation and future conditional variance of the return on a financial asset. Let $\rho<0$ and $\sigma_1<\sigma_2$. In this case, a negative realization of $\epsilon_t$ increases the probability of the second state in period $t+1$ and leads to an increase in the conditional volatility. 

A simple application in time series analysis includes the modeling of the mean reversion that works differently for $\rho$ of different signs. Let $\mu_1<\mu_2$ and $\rho<0$. Then, a positive realization of $\epsilon_t$ decreases the probability of the second state in the period $t+1$. Therefore, the mean reversion is also obtained at the level of the future conditional expected value that now decreases. In the opposite case of $\rho>0$, a positive realization of $\epsilon_t$ increases the probability of the second state in period $t+1$ and, consequently, increases the conditional expected value of $y_t$. Therefore, the forecasts of $y$ revert to a mean that is higher, which has a destabilizing effect. Examples of more elaborate applications in economics include the endogenous switching of the parameters modeling the effects of monetary and fiscal policies proposed by \cite{Chang2018} and \cite{Chang2017a}, respectively.

\section{Future Developments}\label{sec:future}

\noindent Two suggested paths for future developments in the MS and IHMM frameworks include efficient algorithms with the potential for parallel computations and interpretability incurred through sparsity. Improvements on both of these fronts are required to make the analysis of big data sets possible by combining sufficient flexibility of the model with scarce parameterization. 

Existing approaches to MS models rely on the FFBS technique, discussed in Section~\ref{sec:intro}. This filtering and smoothing method consists of an iterative procedure that implies serial computations. With an increasing number of observations and states required to capture the features of data, the FFBS algorithm becomes computationally too requiring for practical implementations. Vectorization, tensor algebra, and efficient numerical algorithms provide some of the solutions. Still, they are far from being as computationally fast as available algorithms for real-values state-space models such as, for instance, the precision sampler by \cite{Chan2009b}.

Moreover, an increasing interest in heterogenous MS models in which individual parameters follow independent Markov processes calls for new methods of allowing sparsity in the modeling. In many such studies, the question of whether time-variation is required for a particular parameter and, if yes, then how many MS regimes are required to model it stay unanswered due to the lack of appropriate algorithms. It is worth mentioning that recent studies provide such solutions for real-values state-space models, and some examples include \cite{Fruhwirth-Schnatter2010}, \cite{Bitto2019}, and \cite{Cadonna2019}.

Similar considerations apply to the IHMM, although it should be emphasized that the IHMM provides certain solutions to the challenges singled out above for the MS models. Here, as the number of observations increases, the number of states, $K$, to be modeled in a particular iteration of the estimation algorithm following the beam sampler step shall increase as well.
This fact increases the computation time required geometrically and calls for a more time-efficient estimation method.
Variational inference offers faster algorithms at the cost of approximating the posterior distribution at an unknown precision \citep[see][for further reference]{wainwright2008graphical}. Variational Bayes captures correctly the central tendency of the approximated distribution. However, it underestimates the posterior variances \citep[see][and references therein]{Wang2019}.
Applications of Variational Bayes estimation to Dirichlet process and IHMM can be found in \cite{blei2006variational}, \cite{kurihara2007collapsed}, \cite{teh2008collapsed}, and \cite{wang2011online}.
Alternative approaches may seek to improve the computation through parallelization such as \cite{fearnhead2004particle}, \cite{rodriguez2011line}, \cite{williamson2013parallel} and \cite{tripuraneni2015particle}.

Notable recent developments granting sparsity in Dirichlet process models were proposed by \cite{fruhwirth2019here}. They build foundations for new developments for the IHMM. In this approach, the sparsity is accompanied by choice of concentration parameters, and as an extra benefit, it offers significantly simplified computations as the sparse structure heavily penalizes the number of states.

\bibliographystyle{model5-names}
\bibliography{ref}

\end{document}